**All-electrical switching of spin texture in a strain-tunable 2D Janus ferroelectric altermagnet**

Tao Yao, Quan Shen, Wenhu Liao, Jianing Tan[a], Jiansheng Dong[a]

Department of Physics, Jishou University, Jishou 416000, Hunan, China

[a]Corresponding authors: jianing_tan@163.com, jsdong@jsu.edu.cn

# Abstract

Altermagnetism (AM), a collinear magnetic phase with momentum-dependent spin splitting, is a promising candidate for strong magnetoelectric coupling. However, realizing direct and tunable coupling between ferroelectricity (FE) and AM within a single two-dimensional (2D) material remains an outstanding challenge. Here, based on first-principles calculations, we identify the distorted phase of monolayer Janus VOClBr as an intrinsic 2D FE-AM. This phase demonstrates robust magnetoelectric coupling, as evidenced by a complete reversal of momentum-space spin polarization upon FE switching, and further supported by spin texture analysis and the magneto-optical Kerr effect. Notably, the FE properties are highly strain-tunable: biaxial compression strain of -4% reduces the FE polarization switching barrier by approximately 87%, whereas a tensile strain of +3% induces a phase transition to an antiferromagnet. Leveraging the lock-in between the electrically controlled spin texture and the magneto-optical Kerr effect signal, we propose a non-volatile, polymorphic spintronic memory device featuring all-electrical writing and optical readout. This work establishes 2D FE-AMs as a versatile platform for coupled ferroic orders and paves the way for voltage-controlled, multifunctional spin-logic devices.

Two-dimensional (2D) multiferroics, which host coupled magnetic and ferroelectric (FE) orders, offer a promising route to ultracompact, low-power electronics by exploiting intrinsic magnetoelectric effects.[1-4] However, the advancement of this class of materials faces two fundamental constraints.[5-7] The first is an electronic incompatibility: ferromagnetism (FM) requires unpaired local moments, whereas conventional FE adheres to the $d^0$ rule, making their coexistence rare. The second is a symmetry constraint: although antiferromagnetic (AFM) multiferroics circumvent this electronic conflict, their inherent symmetries enforce spin degeneracy. This results in ground states that are difficult to probe and control, a limitation that severely restricts their functional utility. These challenges underscore the need for unconventional magnetic orders that are intrinsically compatible with, and controllable through, FE polarization.

The recent discovery of altermagnetism (AM), a third fundamental class of collinear magnetism distinct from FM and AFM, provides a promising avenue.[8-12] AMs are characterized by antiparallel magnetic order in real space that, driven by specific crystalline symmetries, generates anisotropic spin splitting in momentum space. This leads to a unique electronic ground state featuring zero net magnetization alongside non-degenerate, spin-polarized bands. Such a state manifests in emergent phenomena including pronounced nonlinear spin currents,[13] large magnetoresistance,[14] and predicted quantum anomalous Hall states.[15] Crucially, the unique combination of broken time-reversal symmetry and momentum-locked spin polarization in AMs suggests a natural microscopic compatibility with FE order, establishing the conceptual foundation for a new class of multiferroics: FE-AMs.

Recent theoretical advances have delineated design principles for directly coupling FE-AM orders in 2D materials, including monolayers of $VOX_2$, $VSX_2$ (X=Cl, Br, I), and $CrPS_3$.[16,17] The resulting FE-AM coupling represents a mechanism distinct from that in conventional multiferroics, as it arises from a precise symmetry match. Specifically, in these systems, cooperative vanadium displacements, driven by pseudo-Jahn-Teller and Peierls distortions, break the translational symmetry of the magnetic sublattice while simultaneously imposing specific rotational and mirror symmetries.[18,19] This unique symmetry engineering directly enables the coexistence and coupling of the two orders. This spontaneous structural distortion concurrently fulfills the essential criteria for both phenomena: broken inversion symmetry for FE polarization and spin-sublattice inequivalence for AM order.

Parallel to these developments, Janus monolayers, characterized by intrinsic vertical asymmetry, have emerged as a versatile platform for functional material design. Their inherent structural polarity enables tailored functionalities, ranging from giant Rashba spin splitting and efficient photocatalysis to coexisting piezoelectricity and valley polarization.[20-23] These highly tunable properties make them strong candidates for next-generation electronics and spintronics. Thus, discovering novel 2D Janus systems with intrinsic FE-AM order and elucidating their microscopic coupling mechanisms are crucial for advancing the field.

Here, we propose the distorted phase of the Janus monolayer VOClBr as a robust 2D FE-AM. We demonstrate that this phase is both dynamically and thermally stable.

Crucially, it exhibits robust magnetoelectric coupling, as evidenced by the reversible switching of its spin texture and the sign reversal of the magneto-optical Kerr effect (MOKE). Furthermore, its FE switching barrier and electronic properties are highly tunable via biaxial strain. Capitalizing on these attributes, we propose a novel non-volatile spintronic memory device that operates via all-electrical writing and optical readout.

Density functional theory (DFT) calculations were carried out using the Vienna *ab initio* Simulation Package (VASP).[24] The electron-ion interactions were described by the projector-augmented wave (PAW) method, while exchange-correlation effects were treated using the Perdew-Burke-Ernzerhof (PBE) generalized gradient approximation (GGA).[25,26] A plane-wave kinetic-energy cutoff of 600 eV was adopted. The electronic self-consistency criterion was set to $10^{-6}$ eV. A Γ-centred 15×9×1 *k*-point grid was applied to sample the Brillouin zone. A vacuum layer of at least 20 Å was introduced along the out-of-plane direction to eliminate spurious interactions between periodic images. Atomic geometries were optimized until the Hellmann-Feynman forces on all atoms were less than 0.005 eV/Å. The strong on-site Coulomb interactions within the V 3*d* orbitals were accounted for using the GGA+U approach.[16,27] The climbing-image nudged elastic band (CI-NEB) method was employed to map the minimum energy paths and determine the reaction kinetics.[28] The dynamical stability was examined by computing the phonon dispersion spectra using the finite-displacement method as implemented in the PHONOPY code with a 4×4×1 supercell.[29,30] The thermal stability was verified by performing *ab initio* molecular dynamics (AIMD) simulations on a 3×3×1 supercell.[31] The optical conductivity was calculated by employing the Wannier90 package.[32,33]

Monolayer Janus VOClBr crystallizes in the polar space group $P_m$ (No. 6), adopting a distinct Cl-V/O-Br trilayer architecture [Fig. 1(a)]. Its structure is derived from the parent compound $VOCl_2$ through asymmetric halogen substitution.[16] Within this framework, a pseudo-Jahn-Teller distortion, driven by hybridization between O 2*p* and V 3*d* orbitals, triggers a spontaneous off-centering of the V ions along the *x*-axis, thereby establishing a robust intrinsic FE polarization ( $P > 0$ ).[19] Although this FE polarization breaks inversion symmetry, the two AFM sublattices remain connected via a composite $[C_2||t_y]$ symmetry. The preserved translational symmetry $t_y$ enforces spin degeneracy ( $P_S = 0$ ) across the full Brillouin zone, resulting in a conventional FE-AFM phase.

Recent studies indicate that in monolayer $VOX_2$ (X=Cl, Br, I), V-V dimerization driven by the Peierls effect induces slight displacements of the V atoms along the *y*-direction.[16] This lattice distortion defines a distinct structural phase (the distorted phase), and, crucially, is the direct origin of the intriguing magnetoelectric coupling effects observed in this monolayer $VOX_2$ system.[16,17] For monolayer Janus VOClBr, this distorted phase is dynamically and thermally stable [Fig. 1(b)], as evidenced by the absence of imaginary phonon frequencies and by AIMD simulations at 300 K (see the supplementary material, S1). To investigate the chemical bonding characteristics of monolayer Janus VOClBr, the calculated electron localization function reveals that electrons are predominantly localized around the O, Cl, and Br atoms, indicating that the V-O, V-Cl, and V-Br bonds exhibit ionic character (see the supplementary material, S2). In transition-metal strongly correlated systems, the inclusion of different Coulomb repulsion (Hubbard U) parameters may influence their electronic structures. We have calculated the band structures of the monolayer Janus VOClBr with different U values. The results show that observable nonrelativistic spin splitting persists in the band structures across the range of U values considered (see the supplementary material, S3). Furthermore, the magnetic anisotropy energy (MAE) indicates a preference for out-of-plane magnetization (see the supplementary material, S4). Compared to the in-plane magnetization of monolayer $VOI_2$, this reversal of the magnetic easy axis can likely be attributed to the charge transfer driven by electronegativity differences arising from atomic substitution, which alters the orbital hybridization of the V 3d states, thereby tuning the orientation of the magnetic easy axis.[34]

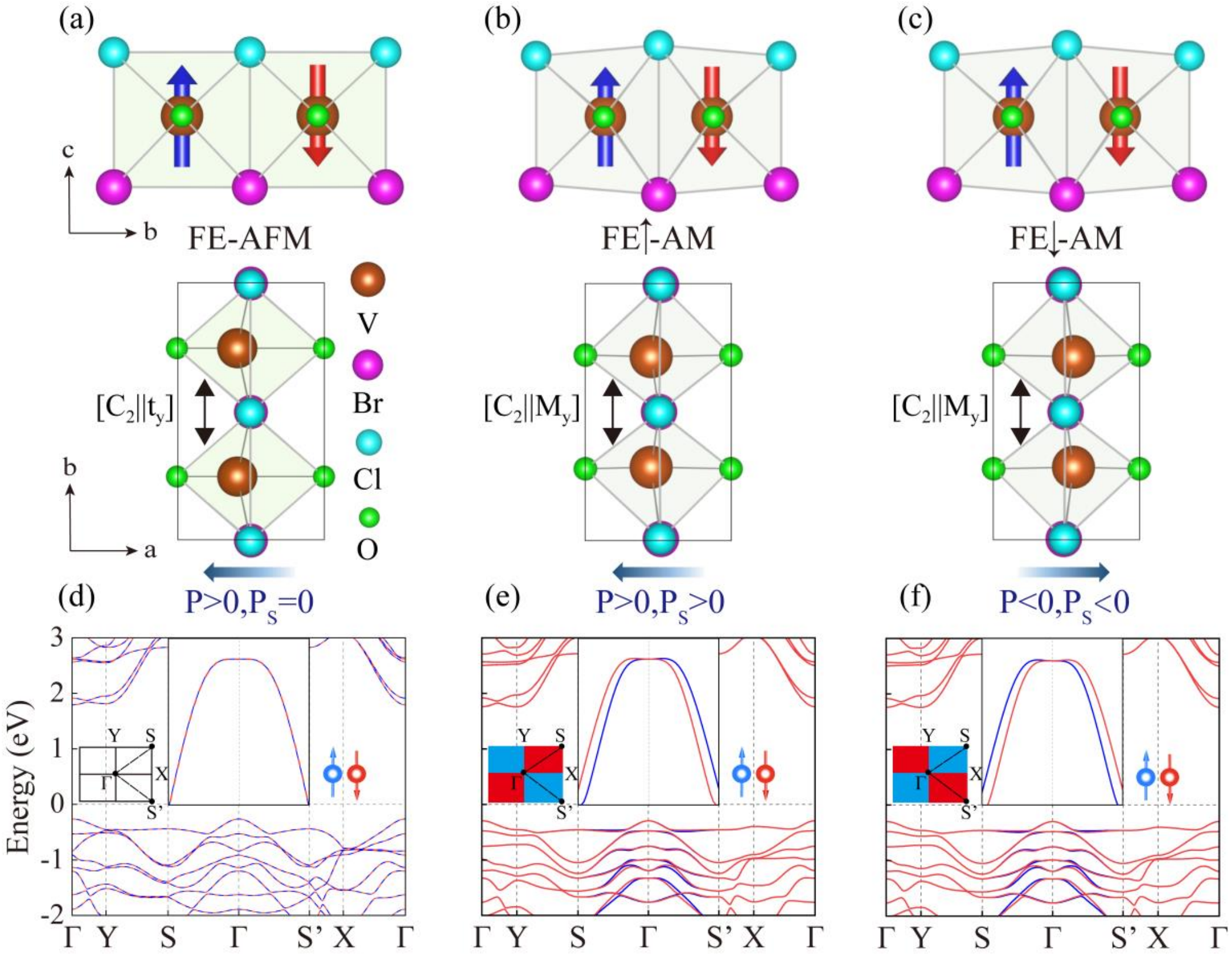

**FIG. 1.** FE-AM in monolayer Janus VOClBr. (a) Structure of the undistorted phase with FE polarization $P > 0$. Distorted configurations for (b) FE↑-AM ( $P > 0$ ) and (c) FE↓-AM ( $P < 0$ ) phases. The red and blue represent opposite-spin magnetic sublattices. (d)-(f) Corresponding spin-polarized band structures for the configurations in (a)-(c), where blue (red) bands indicate spin-up (spin-down) states. The inset shows the Brillouin zone with high-symmetry points.

The distorted phase of monolayer Janus VOClBr exhibits a stable FE polarization. This FE polarization originates from a cooperative off-centering displacement of V atoms, which breaks the spatial inversion symmetry while preserving the mirror symmetry with respect to the *y*-axis. Within this symmetry-broken state, the combined operation of twofold rotation and mirror symmetry (denoted as $[C_2||M_y]$) defines the symmetry that governs the magnetic exchange. Critically, this symmetry forces momentum-dependent spin splitting ($P_S > 0$) along the S-Γ-S' path in the Brillouin zone. This anisotropic spin splitting is a hallmark of AM. Its coexistence with FE establishes a FE↑-AM ground state [Fig. 1(b)] (see the supplementary material, S5, for details of the magnetic ground-state calculations).

For practical device applications, the reversal of FE spontaneous polarization requires a sufficiently low switching barrier (typically below 500 meV).[17] Using the CI-NEB method, we calculated a switching barrier of 390 meV/f.u. for the distorted phase of monolayer Janus VOClBr. Notably, the energy barrier of monolayer $VOI_2$ is 440 meV/f.u.[16] This observation suggests that the intrinsic electronegativity difference induced by the Janus structure influence the orbital hybridization between O 2p and V 3d, thereby partially lowering the energy barrier.

The switching involves displacement of the central V ion between symmetric sites within its octahedron, connecting two degenerate polarization states. This moderate barrier confirms that FE polarization reversal is readily accessible by an applied electric field. Notably, switching the FE polarization in the distorted phase of monolayer Janus VOClBr inverts the momentum-space spin polarization, changing its sign (FE↑-AM↔FE↓-AM) [Fig. 1(c)]. The consequent reversal of the spin-polarized state demonstrates a robust locking between the FE and AM orders.

The magneto-optical Kerr effect (MOKE), a non-contact and sensitive optical probe, directly detects time-reversal symmetry breaking and momentum-dependent spin splitting by measuring polarization rotations in reflected light.[35-37] Consequently, it has emerged as a prominent technique for characterizing magnetic order. Notably, conventional collinear AFMs typically lack a MOKE, as their combined *PT* symmetry enforces spin degeneracy across momentum space. For the distorted phase of monolayer Janus VOClBr, the dielectric function tensor $\varepsilon_{ij}(\omega)$ is directly related to the optical conductivity tensor $\sigma_{ij}(\omega)$ via $\varepsilon_{ij}(\omega)=\delta_{ij}+i\frac{4\pi}{\omega}\sigma_{ij}(\omega)$.[36,38] The symmetry-imposed form of $\sigma_{ij}(\omega)$, dictated by the magnetic space group of the system, can be expressed generally as:[39]

$$\sigma=\begin{bmatrix}\sigma_{xx} & 0 & \sigma_{xz} \\ 0 & \sigma_{yy} & 0 \\ \sigma_{zx} & 0 & \sigma_{zz}\end{bmatrix} \tag{1}$$

Then, the complex Kerr angle $\phi_K$ was calculated using the relation:

$$\phi_K=\theta_K+i\eta_K=\frac{-(\sigma_{xz}-\sigma_{zx})}{2\sigma_{xx}\sqrt{1+\frac{i4\pi}{\omega}\sigma_{xx}}} \tag{2}$$

where $\theta_K$ and $\eta_K$ denote the Kerr rotation and Kerr ellipticity, respectively. The magnetoelectric coupling between FE and spin polarization in the distorted phase of monolayer Janus VOClBr is directly probed by the spin texture and the MOKE. Figure 2 reveals a robust *d*-wave symmetry in in the spin texture, which coherently inverts upon FE polarization reversal [Figs. 2(a) and 2(b)]. This AM order, by breaking time-reversal symmetry, induces a pronounced MOKE through its associated anisotropic optical conductivity [Figs. 2(c) and 2(d)]. Consequently, the measured Kerr rotation ($\theta_K$) and ellipticity ($\eta_K$) provide a direct optical readout of the coupled FE-magnetic state: their signs switch in precise unison with the reversal of both the FE and the AM order. This set of observations establishes a deterministic magnetoelectric coupling in the strict 2D limit.

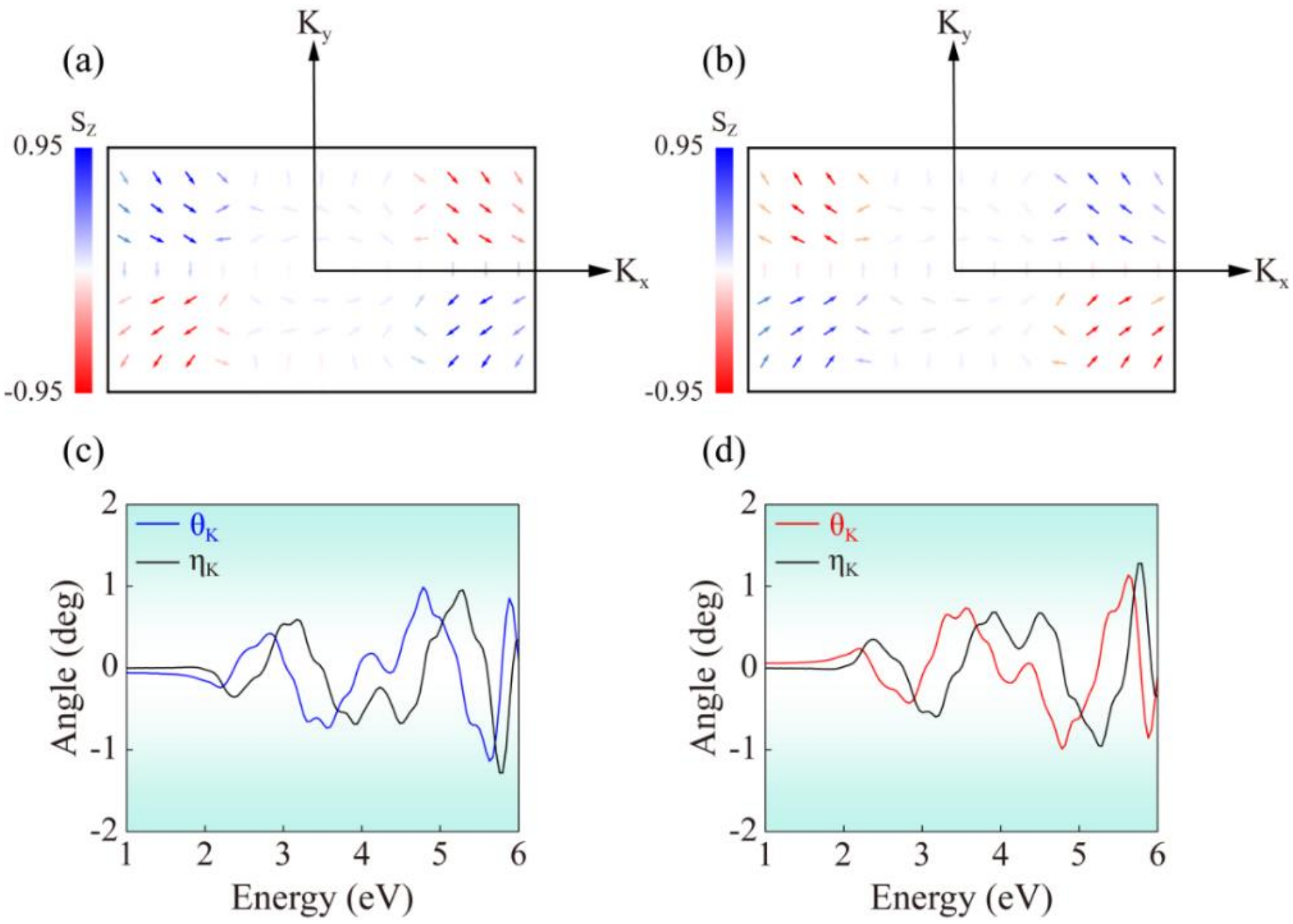

**FIG. 2.** Spin texture and MOKE in the distorted phase of monolayer Janus VOClBr. Calculated in-plane spin textures for the (a) FE↑-AM and (b) FE↓-AM phases. (c), (d) Corresponding magneto-optical Kerr rotation spectra for the phases in (a) and (b), respectively.

Strain engineering provides a powerful strategy for controlling coupled electronic and ferroic orders in 2D materials.[40-42] Using the distorted phase of monolayer Janus VOClBr as a platform, where a Peierls distortion intricately couples FE, magnetism, and electronic structure, we demonstrate how strain can selectively tune these intertwined properties. We investigate the effects of applied biaxial strain ($\varepsilon = \pm 4$ %) on the band structure, magnetic ground state, and FE polarization switching barrier. Under compressive strain, the V-V distance is preserved, which stabilizes the distorted phase. As a result, monolayer Janus VOClBr maintains its FE-AM character, as evidenced by persistent momentum-dependent spin splitting along the S-Γ-S’ path in the Brillouin zone [Figs. 3(a) and 3(b)] (see supplementary material, S6). In contrast, tensile strain suppresses the spin splitting, driving the system toward complete spin degeneracy [Figs. 3(d) and 3(e)]. This transition is caused by the strain-induced increase in the V-V separation along the *y*-direction, which weakens the Peierls distortion and ultimately prevents the electronic dimerization. This structural response directly reshapes the magnetic energy landscape. Specifically, tensile strain progressively diminishes the energy advantage of the AFM1 state over a competing AFM3 phase, ultimately triggering a magnetic phase transition at a critical strain of 3% [Fig. 3(c)].

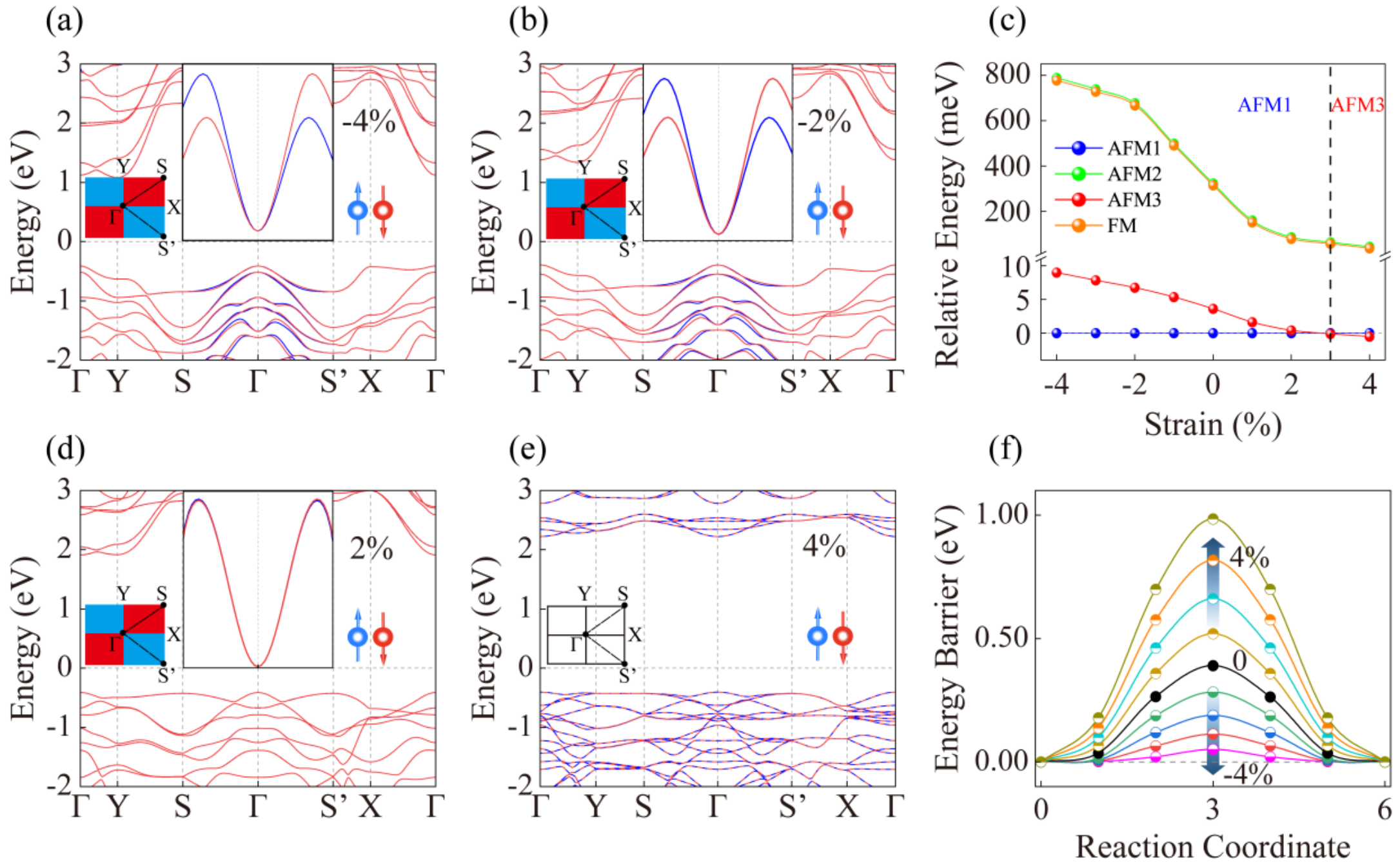

**FIG. 3** Strain engineering of electronic, magnetic, and FE properties in the distorted phase of monolayer Janus VOClBr. Electronic band structures under (a) -4% and (b) -2% compressive biaxial strain. (c) Relative energy of the competing antiferromagnetic states AFM1 and AFM3 as a function of biaxial strain. Band

structures under (d) +2% and (e) +4% tensile biaxial strain. (f) Calculated FE polarization switching energy barrier versus biaxial strain.

Simultaneously, biaxial strain exerts a dramatic and opposing influence on the FE switching barrier [Fig. 3(f)]. Tensile strain ($\varepsilon = +4\%$) increases the barrier by approximately 150% (nearly tripling it), effectively locking the polarization. In stark contrast, compressive strain ($\varepsilon = -4\%$) reduces the barrier to a modest 50 meV, which significantly lowers the energy cost of FE polarization reversal. These results demonstrate that biaxial strain serves as a powerful, multifunctional external knob, capable of controlling magnetic order, electronic structure, and FE switching dynamics in the distorted phase of monolayer Janus VOClBr.

This deterministic magnetoelectric coupling enables a novel device concept that integrates electrical control with optical readout. Building on this concept, we propose a prototype for a non‑volatile memory device. In this device, information is encoded by switching the FE polarization and retrieved via the consequent sign reversal of the MOKE in the FE‑AM phase. An applied voltage switches the FE polarization, thereby inverting both the spin texture and the sign of the Kerr rotation ($\theta_K$) in the reflected light [Fig. 4]. This direct link creates two bistable, non-volatile states: a logic ‘1’ ($P > 0$, $+\theta_K$) and a logic ‘-1’ ($P < 0$, $-\theta_K$) [Figs. 4(a) and 4(b)]. The write process is purely electrical (voltage-driven FE polarization reversal), whereas the readout is all-optical (measurement of $\theta_K$), enabling fast, low-power, and non-destructive operation. It should be emphasized that, this study establishes a universal operational principle that exploits the intrinsic magnetoelectric coupling in FE-AMs, with FE and spin orders serving as the writable and readable states, respectively, via their direct link to optical responses. By combining electrical writing, achieved through FE polarization switching, with optical readout via the MOKE, it provides a generalized platform for the electrical manipulation of spin polarization and magneto-optical responses. This multidimensional coupling enables fast, low-power device operation and holds promise for extension to other ferroelectric-antiferromagnetic (FE-AM) systems with low switching barriers and high Néel temperatures. The approach paves the way toward non-volatile multiferroic memory technologies and highly sensitive opto-spintronic sensors.

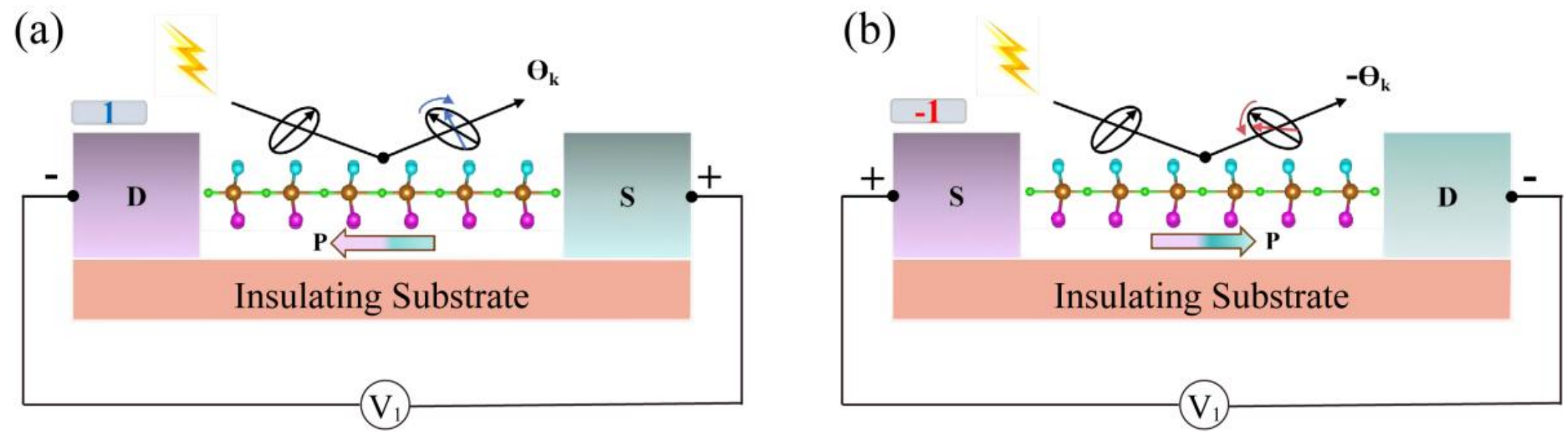

**FIG. 4.** Electrical switching of an all-optical memory state in a FE-AM. (a) FE↑-AM state (logic ‘1’) with FE polarization $P > 0$, where the coupled spin texture yields a positive Kerr rotation ($+\theta_K$) in the reflected light. (b) FE↓-AM state (logic ‘-1’) with FE polarization $P < 0$, where the spin texture is inverted, producing a negative Kerr rotation ($-\theta_K$).

In summary, first-principles calculations identify the distorted phase of monolayer Janus VOClBr as an intrinsic 2D FE-AM. This phase exhibits robust magnetoelectric coupling, whereby the reversal of FE polarization coherently inverts the momentum-space spin texture, enabling full electrical control of the spin polarization. This direct relationship is evidenced by reversible spin-texture flipping and a corresponding sign-switching MOKE, which further enables non-contact optical readout of the FE polarization state. Moreover, biaxial strain provides a powerful external knob to modulate the coupled order: compressive strain significantly reduces the FE switching barrier, facilitating low-power operation, whereas tensile strain quenches the Peierls distortion, inducing a phase transition to a conventional FE-AFM state with improved FE polarization stability. Based on this deterministic "ferroelectric-spin-optical" coupling, we propose a prototype non-volatile, polymorphic spintronic memory device operating via all-electrical writing and optical readout. This study extends the family of 2D multiferroic systems, clarifies the underlying physics of FE-AM coupling, and provides a material platform for multifunctional, energy-efficient spintronic devices that integrate electrical control, spin transport, and optical interfacing.

**Supplementary Material**

See the supplementary material for the dynamically and thermally stable, magnetic configurations, magnetic anisotropy energy, and strain-dependent band structures.

**Acknowledgments**

This work was supported by the National Natural Science Foundation of China (Grant No. 12364007, 12264016), and Scientific Research Foundation of Hunan Provincial Education Department, China (Grant No. 24A0366).

## AUTHOR DECLARATIONS

### Conflict of Interest

The authors have no conflicts to disclose.

## DATA AVAILABILITY

The data that support the findings of this study are available from the corresponding author upon reasonable request.

## REFERENCES

[1] J. W. Chu, Y. Wang, X. Wang, K. Hu, G. Rao, C. Gong, C. Wu, H. Hong, X. Wang, K. Liu, C. Gao, and J. Xiong, Adv. Mater. **33**, 2004469 (2021).

[2] Y. L. Wu, Z. Z. Zeng, H. P. Lu, X. C. Han, C. D. Yang, N. S. Liu, X. X. Zhao, L. Qiao, W. Ji, R. C. Che, L. J. Deng, P. Yan, and B. Peng, Nat. Commun. **15**, 8616 (2024).

[3] J. T. Jiang, F. Wu, Y. Wan, A. Li, C. Huang, and E. Kan, Phys. Rev. Lett. **134**, 196801 (2025).

[4] H. J. Wang, H. T. Liu, M. Ye, and Y. C. Li, Phys. Rev. Lett. **135**, 226402 (2025).

[5] L. Šmejkal, Y. Mokrousov, B. Yan, and A. H. MacDonald, Nat. Phys. **14**, 242 (2018).

[6] S. Dong, H. J. Xiang, and E. Dagotto, Natl. Sci. Rev. **6**, 629 (2019).

[7] H. Y. Chen, L. Liu, X. R. Zhou, Z. Meng, X. N. Wang, Z. Y. Duan, G. J. Zhao, H. Yan, P. X. Qin, and Z. Q. Liu, Adv. Mater. **36**, 2310379 (2024)

[8] L. Šmejkal, J. Sinova, and T. Jungwirth, Phys. Rev. X. **12**, 040501 (2022).

[9] Q. Shen, W. H. Liao, H. R. Bao, D. G. Xu, J. N. Tan, J. S. Dong, and G. Ouyang, Phys. Rev. Mater. **9**, 114404 (2025).

[10] Y. C. She, Y. D. Wang, H. B. Sun, C. Wu, W. X. Zhang, and P. Li, Phys. Rev. B. **113**, 035420 (2026).

[11]S. A. A. Ghorashi, T. L. Hughes, and J. Cano, Phys. Rev. Lett. **133**, 106601 (2024).

[12] S. D. Guo, X. S. Guo, and G. Wang, Phys. Rev. B. **110**, 184408 (2024).

[13]R. G. Hernández, L. Šmejkal, K. Výborný, Y. Yahagi, J. Sinova, T. Jungwirth, and J. Železný, Phys. Rev. Lett. **126**, 127701 (2021).

[14]L. Šmejkal, A. B. Hellenes, R. G. Hernández, J. Sinova, and T. Jungwirth, Phys. Rev. X. **12**, 011028 (2022).

[15]P. J. Guo, Z. X. Liu, and Z. Y. Lu, npj Comput. Mater. **9**, 1 (2023).

[16]Z. Y. Zhu, X. K. Duan, J. Y. Zhang, B. W. Hao, I. Žutić, and T. Zhou, Nano Lett. **25**, 9456 (2025).

[17]S. Y. Wang, W. W. Wang, J. X. Fan, X. D. Zhou, X. P. Li, and L. Wang, Nano Lett. **25**, 14618 (2025).

[18]Y. Zhang, L. F. Lin, A. Moreo, G. Alvarez, and E. Dagotto, Phys. Rev. B. **103**, L121114 (2021).

[19]H. X. Tan, M. L. Li, H. T. Liu, Z. R. Liu, Y. C. Li, and W. H. Duan, Phys. Rev. B. **99**, 195434 (2019).

[20]T. Hu, F. H. Jia, G. D. Zhao, J. Y. Wu, A. Stroppa, and W. Ren, Phys. Rev. B. **97**, 235404 (2018).

[21]J. Zhao, Y. X. Qi, C. Yao, and H. Zeng, Phys. Rev. B. **109**, 035408 (2024).

[22]D. Q. Er, H. Ye, N. C. Frey, H. Kumar, J. Lou, and V. B. Shenoy, Nano Lett. **18**, 3943 (2018).

[23]L. Dong, J. Lou, and V. B. Shenoy, ACS Nano. **11**, 8242 (2017).

[24]G. Kresse and J. Furthmüller, Phys. Rev. B. **54**, 11169 (1996).

[25]J. P. Perdew, K. Burke, and M. Ernzerhof, Phys. Rev. Lett. **77**, 3865 (1996).

[26]P. E. Blöchl, Phys. Rev. B. **50**, 17953 (1994).

[27]S. L. Dudarev, G. A. Botton, S. Y. Savrasov, C. J. Humphreys, and A. P. Sutton,

Phys. Rev. B. **57**, 1505 (1998).

[28]G. Henkelman, B. P. Uberuaga, and H. Jónsson, J. Chem. Phys. **113**, 9901 (2000).

[29]A. Togo, J. Phys. Soc. Jpn. **92**, 012001 (2023).

[30]A. Togo, L. Chaput, T. Tadano, and I. Tanaka, J. Phys.: Condens. Matter. **35**, 353001 (2023).

[31]G. Kresse and J. Hafner, Phys. Rev. B. **49**, 14251 (1994).

[32]G. Pizzi, V. Vitale, R. Arita, S. Blügel, F. Freimuth, G. Géranton, M. Gibertini, D. Gresch, C. Johnson, T. Koretsune, J. Phys.: Condens. Mat-ter **32**, 165902 (2020).

[33]I. Souza, N. Marzari, D. Vanderbilt, Phys. Rev. B **65**, 035109 (2001).

[34]Z. J. Zhao, Z. N. Sun, X. R. Li, Y. J. Yu, Physica. B **670**, 415369 (2023).

[35]T. Higo, H. Man, D. B. Gopman, L. Wu, T. Koretsune, O. M. J. van't Erve, Y. P. Kabanov, D. Rees, Y. Li, M. T. Suzuki, S. Patankar, M. Ikhlas, C. L. Chien, R. Arita, R. D. Shull, J. Orenstein, and S. Nakatsuji, Nat. Photon. **12**, 73 (2018).

[36]K. Yang, W. T. Hu, H. Wu, M. H. Whangbo, P. G. Radaelli, and A. Stroppa, ACS Appl. Electron. Mater. **2**, 1373 (2020).

[37]N. Ding, K. Yananose, C. Rizza, F. R. Fan, S. Dong, and A. Stroppa, ACS Appl. Mater. Interfaces. **15**, 22282 (2023).

[38]D. Sangalli, A. Marini, and A. Debernardi, Phys. Rev. B. **86**, 125139 (2012).

[39]M. Born and E. Wolf, Principles of Optics: Electromagnetic Theory of Propagation, Interference and Diffraction of Light (Elsevier, Amsterdam, 2013).

[40]Q. Ma, B. Wang, G. Yang, and Y. Liu, Appl. Phys. Lett. **126**, 223106 (2025).

[41]S. D. Guo and G. Z. Wang, Appl. Phys. Lett. **127**, 042402 (2025).

[42]Q. R. Cui, J. H. Liang, Z. J. Shao, P. Cui, and H. X. Yang, Phys. Rev. B. **102**, 094425 (2020).